\documentclass[10pt,letterpaper]{article}
\usepackage{opex3}
\usepackage{amssymb}
\pdfoutput=1
\begin{document}

\title{Analysis of Kapitza-Dirac diffraction patterns beyond the Raman-Nath regime }

\author{Bryce Gadway, Daniel Pertot, Ren{\'e} Reimann, Martin G. Cohen, and Dominik Schneble}

\address{Department of Physics and Astronomy, Stony Brook University, \\ Stony Brook, NY 11794-3800}

\email{bgadway@ic.sunysb.edu}



\begin{abstract*} We study Kapitza-Dirac diffraction of a Bose-Einstein condensate from a standing light wave for a square pulse with variable pulse length but constant pulse area. We find that for sufficiently weak pulses, the usual analytical short-pulse prediction for the Raman-Nath regime continues to hold for longer times, albeit with a reduction of the apparent modulation depth of the standing wave. We quantitatively relate this effect to the Fourier width of the pulse, and draw analogies to the Rabi dynamics of a coupled two-state system. Our findings, combined with numerical modeling for stronger pulses, are of practical interest for the calibration of optical lattices in ultracold atomic systems.\end{abstract*}


\bibliographystyle{osajnl}

\section{Introduction}
The diffraction of matter-waves from a standing light wave is a fundamental concept in atom optics \cite{Adams-94,Meystre}. Originally predicted by Kapitza and Dirac \cite{Kapitza-33} for electrons more than 75 years ago (and recently also observed for these \cite{Freimund-02}), it was first demonstrated in the 1980s with an atomic beam \cite{Gould-86}, and has since become a standard tool in atom interferometry for the coherent mixing of momentum modes \cite{Berman-97,Meystre}. The advent of Bose-Einstein condensates in the 1990s~\cite{VarennaBEC} has made it possible to directly observe the dynamics of matter-wave diffraction in time-of-flight images \cite{Ovchinnikov-99,Denschlag-02,Huckans-09}. Moreover, the diffraction of condensate atoms from standing light waves finds applications in high-resolution spectroscopy \cite{Stenger-99} and metrology \cite{Gupta-02,Campbell-05}, and plays a fundamental role in superradiance \cite{Inouye-99,Schneble-03}.

Two diffraction scenarios can generally be distinguished. In the resonant Bragg case, the atoms oscillate between two resonantly coupled plane-wave momentum states depending on the strength and duration of the interaction with the light field~\cite{Martin-88}. If the Bragg condition is not met, atoms can nevertheless be diffracted into a number of off-resonant momentum states, provided that the interaction is sufficiently short and strong~\cite{Gould-86}. In accordance with a common convention in atom optics~\cite{Gupta-01} we refer to this case as Kapitza-Dirac diffraction, but we extend it to include times beyond the Raman-Nath regime. In this context, it is interesting to note a similar discussion for the diffraction of light from sound waves in acousto-optic devices \cite{Klein-67}.

In the Raman-Nath regime, i.e. when atomic motion during the interaction with the light field can be neglected, the populations of the diffracted states exclusively depend on the product of the strength $V_0$ and the duration $\tau$ of the interaction, i.e. the area of the applied pulse~\cite{Gould-86}. Outside the Raman-Nath regime, the diffraction dynamics exhibits collapses and revivals for constant interaction strength~\cite{Ovchinnikov-99,Denschlag-02,Huckans-09}. In this paper, we now specifically analyze the case of a pulsed interaction of variable duration but constant pulse area. This allows for a study of the breakdown of the Raman-Nath prediction and, in particular, for a quantification of deviations when the system is close to, but not deep into, the Raman-Nath regime.

The findings of our study are of direct interest for the elimination of systematic errors in calibration measurements for experiments with ultracold atoms in optical lattices~\cite{Morsch-06,Bloch-08} when the lattice depth is determined via Kapitza-Dirac diffraction.

This paper is organized as follows: Section \ref{SEC:Raman-Nath} reviews general aspects of the matter-wave diffraction from a one-dimensional optical lattice, while Section \ref{SEC:ExpProc} briefly describes our experimental system. In Section \ref{SEC:WeakPulse} we analyze diffraction patterns for weak pulses, and give an analytical modification to the Raman-Nath diffraction formula, which is motivated by a comparison of the diffraction dynamics with that of a coupled Rabi system, as well as by considering the spectral properties of the pulse. Section \ref{SEC:StrongPulse} discusses the more complicated dynamics of strong pulses, and Section \ref{SEC:OLCalibration} deals with calibration methods for the depth of optical lattices, comparing numerical simulations with single-shot diffraction patterns not restricted to the Raman-Nath regime.

\section{Raman-Nath regime}
\label{SEC:Raman-Nath}
We first briefly review general aspects of the diffraction of a condensate from a standing light wave with wavenumber $k=2\pi/\lambda$ that is switched on for a duration $\tau$. The standing wave gives rise to a sinusoidal optical potential $V_0 \cos^2 k z$~\cite{Morsch-06}. The evolution of the condensate in the standing wave can then be modeled ~\cite{Cook-78,Batelaan-07} (neglecting mean-field interactions) as that of a matter wave $\psi$ subject to the Hamiltonian
\begin{equation}
\hat{H} = - (\hbar^2/2 m)\partial_z^2 + V_0 \cos^2 k z
\label{EQ:Hamiltonian}
\end{equation}
where $m$ is the atomic mass. Expanding the condensate wave function in the basis of plane waves populated by diffraction from a standing wave as $\psi(t) = \sum_n c_n(t)~ e^{i 2 n k z}$ (where $n = 0, \pm 1, \pm 2,\dots$ and $c_n(t=0) = \delta_{n,0}$) and introducing the dimensionless parameters
\begin{eqnarray}
\alpha & =&  (E_r^{(2)}/\hbar)~\tau\\
\beta & =&  (V_0/\hbar) ~\tau,
\end{eqnarray}
(where $E_r^{(n)} = (n\hbar k)^2/2m$ denotes the $n$-photon recoil energy, with $E_r^{(1)}\equiv E_r$), transforms the time-dependent Schr{\"o}dinger equation into a set of coupled differential equations
\begin{equation}
i \frac{\mbox{d} {c}_n}{\mbox{d} t} = \frac{\alpha~n^2}{\tau}~c_n + \frac{\beta}{4 \tau}\left(c_{n-1} + 2 c_n + c_{n+1}\right)
\label{EQ:coupled1}
\end{equation}
for the amplitudes $c_n(t)$ of the diffracted orders $n$. For a given lattice depth, the highest momentum order $(\pm 2 n \hbar k$) capable of being populated is given by the cutoff
\begin{equation}
\bar{n} = \sqrt{\beta/\alpha}
\end{equation}
for which the potential energy is fully converted into kinetic energy (cf. \cite{Huckans-09}). Note that in these equations, $\alpha$ is the pulse duration $\tau$ in units of the 2-photon recoil time $\tau_r^{(2)} = \hbar/E_r^{(2)}$, and $\beta$ measures the area of the pulse.

The dynamics of the condensate in the standing wave depends on the ratio between $\alpha n^2$ and $\beta$, i.e. between the kinetic energies acquired during diffraction and the depth of the potential. The Raman-Nath approximation consists of neglecting the $\alpha n^2$ terms in Eqs.~(\ref{EQ:coupled1}), which is justified if $\tau$ is much shorter than the harmonic oscillation period in a potential well~\cite{Ovchinnikov-99}, such that $\tau\omega_{ho}\ll 1$, with $\omega_{ho} = [V_0 E_r^{(2)}]^{1/2}/\hbar$, or equivalently $\beta\alpha\ll 1$. In this case, the solution of Eqs.~(\ref{EQ:coupled1}) is $c_n(t) = (-i)^n e^{-i \beta t/2\tau} J_n(\beta t/2\tau)$, such that the population $P_n = |c_n|^2$ of the $n$th diffracted order after application of the pulse is given by
\begin{equation}
P_n = J_n^2\left(\frac{\beta}{2}\right),
\label{EQ:tradKD}
\end{equation}
where the $J_n$ are Bessel functions of the first kind.

\section{Experimental procedure}
\label{SEC:ExpProc}
In the experiments described in this paper, we subject an optically trapped $^{87}$Rb Bose-Einstein condensate to a vertically-oriented, pulsed standing light wave at 1064~nm, for which the 2-photon recoil time $\tau_r^{(2)}\approx 20~\mu$s. A description of our apparatus for the production of condensates is given in ref.~\cite{Pertot-09a}. In brief, in a nearly isotropic crossed-beam optical dipole trap we produce condensates typically containing $5\times10^5$ atoms in the $|F=1,m_F=-1\rangle$ hyperfine ground state without a discernible thermal fraction.  The standing light wave, with a Gaussian $1/e^2$ radius of $130~\mu$m at the position of the condensate, is derived from a single-frequency ytterbium fiber laser with a linewidth of 70~kHz and can be switched off within $1.5~\mu$s using an acousto-optic modulator. Immediately after application of the pulse, the atoms are released from the optical trap and are imaged after 15~ms time-of-flight using near-resonant absorption imaging on the cycling transition.

\begin{figure*}[h!!tbp]
  \centering\includegraphics[width=\columnwidth]{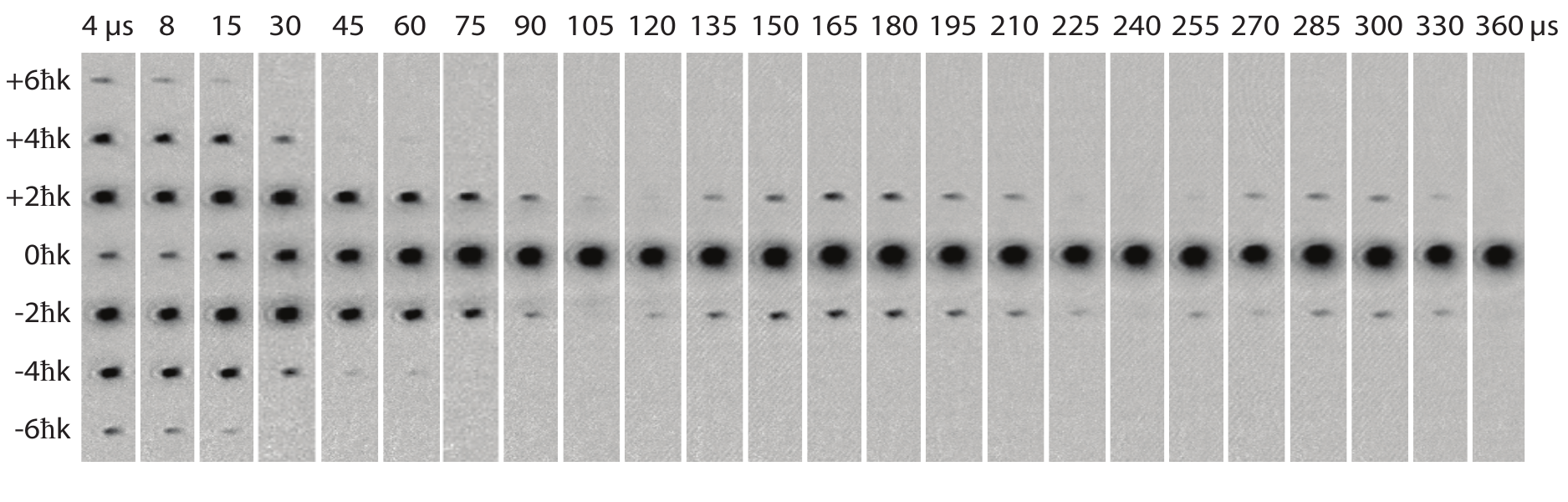}
    \caption{Time-of-flight (TOF) absorption images of a condensate diffracted from a 1064 nm standing-wave optical pulses of constant area $\beta=V_0 \tau/\hbar=4.5$, and durations ranging from $\tau = 4~\mu$s ($\alpha = \tau/\tau_r^{(2)}\thickapprox 0.2$) to 360~$\mu$s ($\alpha \thickapprox18$). For durations $\tau \gtrsim  75\ \mu s$ (for which $[\beta/\alpha]^{1/2}\lesssim 1.1$), the apparent modulation depth of the standing wave undergoes an oscillatory decay consistent with the form $(\beta/2)\mbox{sinc}(\alpha/2)$  (see text).}
  \label{FIG:MiniTreePlot}
\end{figure*}

A series of images taken for pulses of variable duration $\tau$ but with constant pulse area $\beta=4.5$ is shown in Fig.~\ref{FIG:MiniTreePlot}. Based on a naive invocation of Eq.~(\ref{EQ:tradKD}) for the Raman-Nath regime one would expect the same diffraction pattern in all the images. Instead, the number of diffracted orders is seen to generally decrease with the duration of the pulse, with an oscillatory decay of the first diffracted orders for long pulse durations.

The populations of the $0\hbar k$ and $\pm 2\hbar k$ momentum orders in the series of Fig. \ref{FIG:MiniTreePlot} are shown in Figs. \ref{FIG:SincDecay} (A) and (B), together with corresponding data for a weaker standing wave with $\beta = 1.5$, for which diffracted orders $2 n\hbar k$ with $|n|>1$ are not populated over the full range of pulse durations. For the stronger pulse with $\beta = 4.5$, this is only the case for pulse durations exceeding $75~\mu$s due to contributions of the $\pm 4\hbar k$ orders at shorter durations, as shown in inset~(C).

\section{Weak-pulse dynamics}
\label{SEC:WeakPulse}

\begin{figure}[t!]
  \centering\includegraphics[width=9cm]{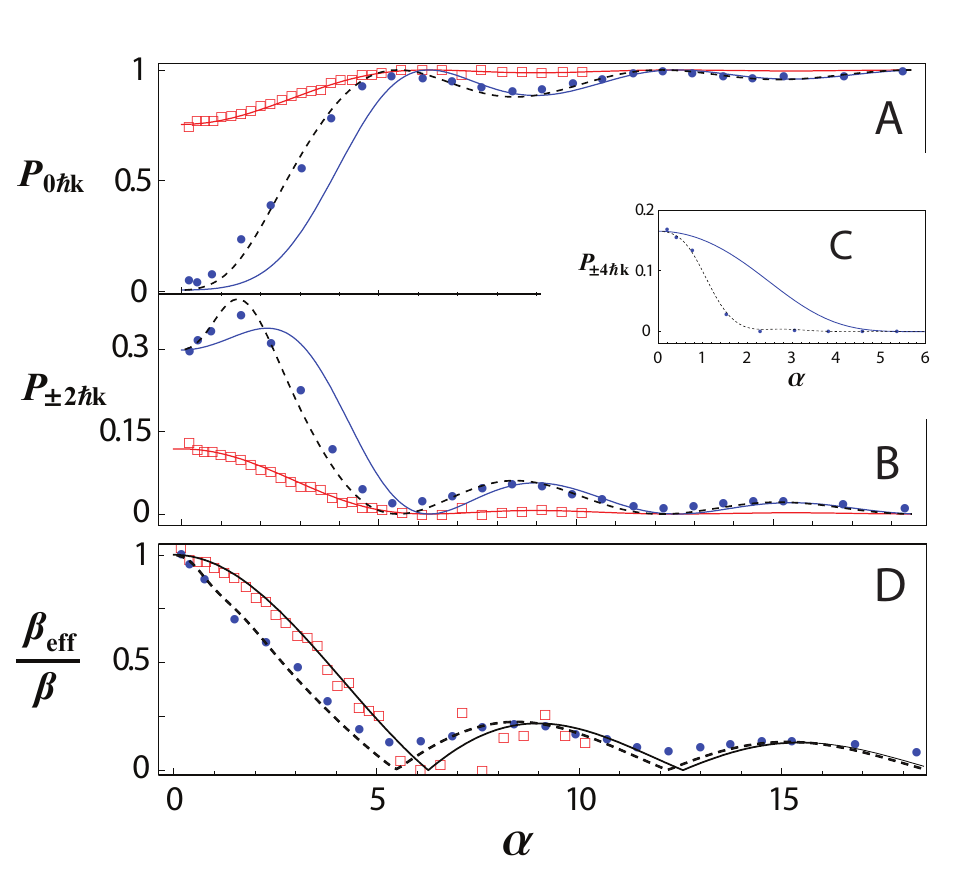}
    \caption{Suppression and revival of atomic diffraction from a constant-area standing wave light pulse with increasing pulse length.  (A and B) Relative populations of the central condensate $0\hbar k$ and orders $\pm2 \hbar k$ vs. normalized pulse duration $\alpha=\tau/\tau_r^{(2)}\sim \tau/20~\mu$s.  Filled blue dots and open red squares refer to pulses of area $\beta=4.5$ (cf.\~Fig. 1) and 1.5, respectively, and the solid lines plot the function $J_n^2[(\beta/2)\mbox{sinc}(\alpha/2)]$.  The dashed black lines correspond to numerical fits for $\beta=4.5$.  (C) Decay of the $\pm 4\hbar k$ orders in the $\beta=4.5$ data set, with the numerical simulation (dashed line) as well as the function $J_2^2[(\beta/2)\mbox{sinc}(\alpha/2)]$.  (D) Behavior of $\beta_{\mbox{\tiny eff}}/\beta$, where the atomic diffraction patterns are fit with the distribution $J_n^2(\beta_{\mbox{\tiny eff}}/2)$.  The black solid line is $\mbox{sinc}(\alpha/2)$, and the dashed black line corresponds to a numerical fit using $\beta=4.5$.}
    \label{FIG:SincDecay}
\end{figure}

For sufficiently weak pulses with a cutoff $\bar{n}\approx1$ for which only the lowest orders $\pm2\hbar k$ are populated, Eqs.~(\ref{EQ:coupled1}) can be reduced to three coupled equations; this approximation is valid for shallow potentials $V_0 \lesssim 4 E_r$. The solutions $c_0$ and $c_{\pm1}$ to Eqs.~(\ref{EQ:coupled1}), which can be obtained in a straightforward way, then lead to
\begin{equation}
P_{\pm 1}  =    \frac{\beta^2}{2\beta^2 +4\alpha^2}~\sin^2\left(\frac{\sqrt{\beta^2/2 +\alpha^2}}{2}~\right)
\label{EQ:Pendel}
\end{equation}
and $P_0 = 1 - 2 P_{\pm1}$. This can be cast into the standard form of a Rabi oscillation
\begin{equation}
P_{\pm 1} = \frac{1}{2}\left(\frac{\chi}{\Omega}\right)^2 \sin^2\left(\frac{\Omega}{2}\tau\right)
\label{EQ:Pendel2}
\end{equation}
with generalized Rabi frequency $\Omega = \left[\chi^2 + \Delta^2\right]^{1/2}$, resonant coupling $\chi = V_0/\sqrt{2}\hbar$ and detuning $\Delta = E_r^{(2)}/\hbar = \omega_r^{(2)}$ between the atomic momentum states $0\hbar k$ and $\pm2\hbar k$. Another way to interpret this result is the following: each scattering event that changes the atomic momentum from $0$ to $\pm 2\hbar k$ by energy conservation leads to a frequency mismatch of the scattered photon by $\Delta$ with respect to the standing wave, requiring a corresponding Fourier width of the pulse for stimulated scattering to be able to occur. Consequently, the drop-off of populations in higher orders with increasing $\alpha$ can be seen as resulting from the decrease in the Fourier width of the pulse, which motivates a more general modification of Eq.~(\ref{EQ:tradKD}) to
\begin{equation}
P_n = J_n^2\left(\frac{\beta}{2}~\mbox{sinc}\frac{\alpha}{2}\right),
\label{EQ:sincFit}
\end{equation}
where the $\mbox{sinc}$ function arises from the Fourier transform of the square pulse. Indeed, the populations $P_{0,\pm1}$ predicted by Eq.~(\ref{EQ:sincFit}) agree with those of Eq. (\ref{EQ:Pendel}) up to $O(\alpha^2\beta^2)$, independent of the pulse duration $\alpha$ or pulse area $\beta$. For $[\beta/\alpha]^{1/2} \leq 1$, the agreement of Eq. (\ref{EQ:sincFit}) with the experimentally observed patterns is excellent, as can be seen in Fig. \ref{FIG:SincDecay} (A) and (B).

It is interesting to note the connection of Eq.~(\ref{EQ:sincFit}) to the resonant case of $n$th order Bragg diffraction from a moving standing wave (with a relative detuning $\delta\omega = n^2\Delta$ between the two beams). In this case, the Fourier transform of the applied potential $V_0 \cos^2 (k z + \delta\omega~t/2)$  with duration $\tau$ would lead to a frequency distribution $\propto$ sinc$[(\omega - \delta\omega) \tau/2]$ (i.e. identical to that of the standing wave case, but shifted by $\delta\omega$) making available the resonant frequency component, independent of the pulse duration.

The form of Eq. (\ref{EQ:sincFit}) shows that although the populations $P_n$ in a spectrum may follow a Bessel distribution as predicted by the Raman-Nath Eq. (\ref{EQ:tradKD}), a fit will not necessarily return the correct modulation depth. The apparent pulse area in Eq. (\ref{EQ:sincFit}) is given by $(\beta/2)\mbox{sinc}(\alpha/2)$, which means that for a pulse of duration $\tau$, the apparent modulation depth of the standing wave extracted from a diffraction spectrum is reduced from $V_0$ to
\begin{equation}
V_{0,\mbox{\footnotesize eff}} = V_0~\mbox{sinc}\left(\tau/2\tau_r^{(2)}\right).\
\label{EQ:Veff}
\end{equation}
The apparent pulse area returned by a fit of Eq. (\ref{EQ:tradKD}) to the diffraction patterns is shown in Fig. \ref{FIG:SincDecay} (D) for both the $\beta = 4.5$ and  $\beta=1.5$ data sets. For the latter, the fit results agree with Eq. (\ref{EQ:Veff}) over the full range, consistent with the fact that orders $|n|>1$ are not populated. The first-order revivals can thus be interpreted as arising from a sampling of the Fourier transform of the pulse. For $\beta=4.5$, the observed behavior still shows qualitative agreement with revivals from sampling of the sinc-shaped frequency distribution with increasing pulse duration $\tau$, but clearly exhibits deviations when higher-order momentum states are present.

\section{Strong-pulse dynamics}
\label{SEC:StrongPulse}
For more intense pulses with $\bar{n} > 1$, the simple generalization of Eq. (\ref{EQ:sincFit}) no longer holds, as is evident by the large discrepancies in Fig. \ref{FIG:SincDecay} for $\tau<75~\mu$s in the $\beta=4.5$ data set.

The presence of higher diffraction orders leads to population decays that are faster than those determined by the two-photon timescale in Eqs.~(\ref{EQ:sincFit}) and (\ref{EQ:Veff}). This is shown in inset (C) of Fig. \ref{FIG:SincDecay} for the $\pm 4 \hbar k$ ($n=2$) orders, which require the exchange of four photons in the transition from $0\hbar k$. In general, the presence of many higher momentum orders will complicate analytic descriptions of the diffraction dynamics.

Nevertheless, it is possible to accurately describe the observed dynamics by numerical integration of the coupled differential Eqs. (\ref{EQ:coupled1}), truncated for unpopulated higher orders beyond $\bar{n}$. This is shown in Figs. \ref{FIG:SincDecay} and \ref{FIG:VaryDepthPlot} (A).

\section{Depth calibration of optical lattices}
\label{SEC:OLCalibration}

In experiments with ultracold atoms in optical lattices, the tunneling rate depends exponentially on the lattice depth $V_0$~\cite{Morsch-06}, for which an accurate knowledge thus is essential. Unlike most other methods (cf. ~\cite{Morsch-06}), applying  a ``short'' Kapitza-Dirac diffraction pulse can conveniently reveal the lattice depth in a single-shot measurement. However, the application of well-defined pulses that are short enough to be deeply in the Raman-Nath regime, yet strong enough to yield significant diffraction, can be technically challenging. The present study quantifies how the results of such a calibration measurement need to be corrected if pulses of finite length are used.

\begin{figure}[t!]
    \centering\includegraphics[width=.9\columnwidth]{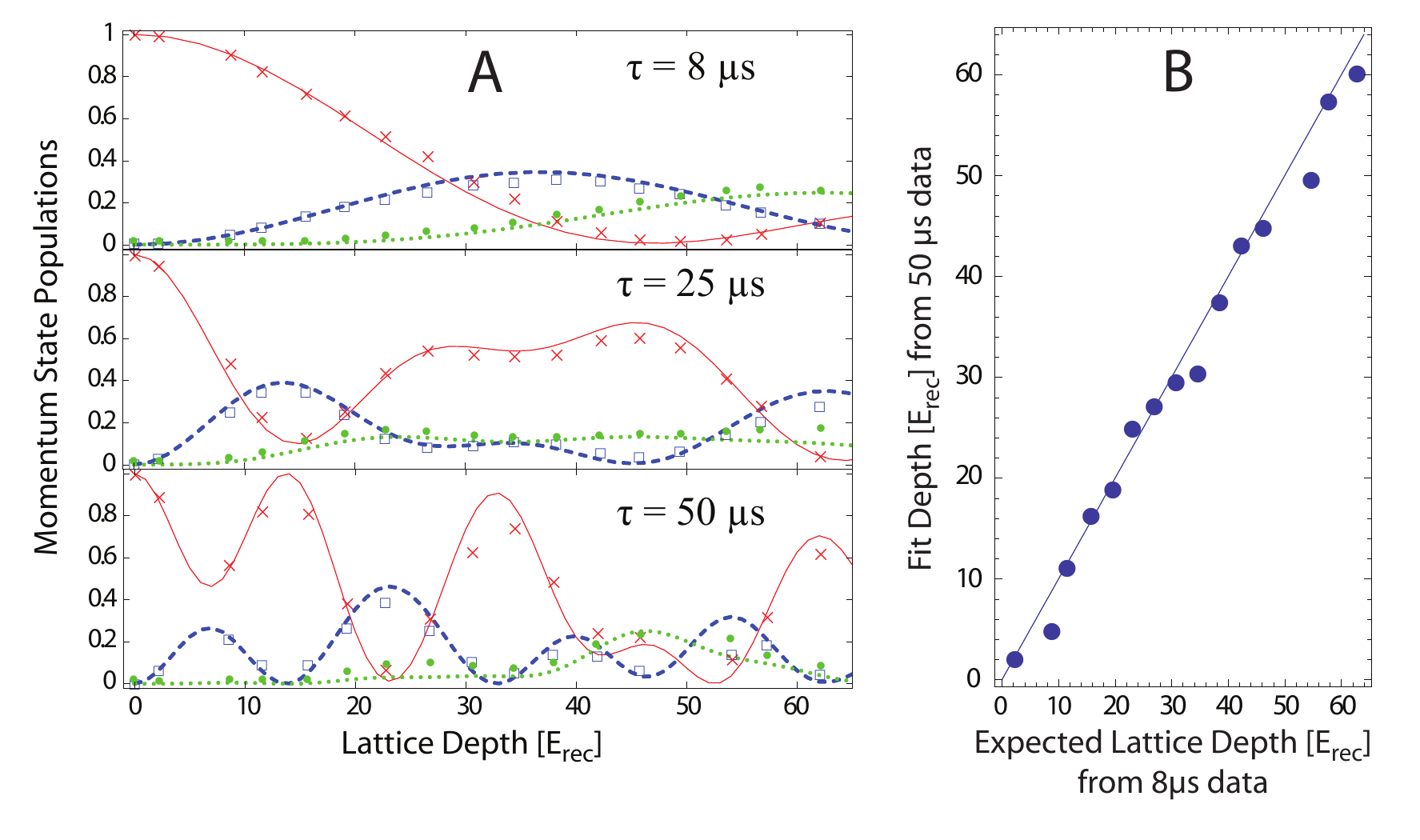}
    \caption{(A) Breakdown of the analytical description.  For each data set, numerical simulations (using Eqs. (\ref{EQ:coupled1})) of the momentum state distribution are shown as solid curves.  The red crosses are data for the $0\hbar k$ order, and the open squares (filled dots) are data averaged for the $\pm2\hbar k$ ($\pm 4\hbar k$) orders. (B) Comparison of lattice depth calibrations from relatively short (8 $\mu$s) and long (50 $\mu$s) optical lattice pulses for various intensities of the standing light wave.  The results for numerically-fit lattice depths agree to within 4\% (the solid line has a slope of 1).}
   \label{FIG:VaryDepthPlot}
\end{figure}

In Fig. \ref{FIG:VaryDepthPlot} we present the comparison of experimentally observed momentum state distributions to numerical simulations, for three different pulse lengths (8, 25, and 50 $\mu$s), as the lattice depth is varied from 0 to 65 $E_r$.  While the observed momentum state distributions are close to the typical Raman-Nath form Eq.~(\ref{EQ:tradKD}) for the 8 $\mu$s pulses, this is no longer the case for longer pulses. For all pulse lengths, numerical simulations agree well, as can be seen in Fig.~\ref{FIG:VaryDepthPlot}~(A).  The lattice depth calibrations for the three pulse durations, determined by comparison of experimental data to numerically simulated patterns, agree to within 4\%, as can be seen in Fig.~\ref{FIG:VaryDepthPlot}~(B) for the extreme cases of 8 and 50 $\mu$s.  Despite the complicated form the dynamics take, one can thus obtain a reliable lattice depth calibration even with longer pulse durations. We have independently verified the method described here with another single-shot calibration method for deep lattices (adiabatic lattice rampup, followed by a sudden projection of the ground band population onto the $\pm 2 \hbar k$ plane-wave states \cite{Christiani-02}), for which we found comparable agreement to within 10\%.

\section{Conclusion}
We have studied Kapitza-Dirac diffraction of a Bose-Einstein condensate in a pulsed standing light wave, considering the case that the pulse area remains constant as the pulse duration is varied. We find that for sufficiently weak pulses exciting only the $\pm 2 \hbar k$ orders, the usual analytical short-pulse prediction continues to hold for longer times, albeit with a modification of the apparent modulation depth of the standing wave. We relate this effect to the frequency spread of the square-wave pulse, and also draw analogies to the Rabi dynamics of a coupled two-state system. Our findings are of practical interest for the calibration of optical lattices in ultracold atomic systems, and we show that for a general length and strength pulse, relatively simple (neglecting mean-field effects) numerical modeling can be used well outside the Raman-Nath regime to accurately determine the lattice depth.

\section*{Acknowledgments}
This work was supported by NSF (PHY-0855643), ONR (DURIP), The Research Foundation of SUNY, and a GAANN fellowship from the U.S. Department of Education (B.G.)
\end{document}